\documentclass[12pt]{article}
\usepackage{natbib,graphicx,fullpage,doublespace,amssymb,amsfonts}
\begin{document}
\bibliographystyle{hmm}

\centerline{\large\bf STATISTICAL PHYSICS OF RUPTURE IN HETEROGENEOUS MEDIA}
\vspace{1em}

The damage and fracture of materials are technologically of enormous
interest due to their economic and human cost. They cover a wide range of 
phenomena like e.g. cracking of glass, aging of concrete, the failure of fiber
networks in the formation of paper and the breaking of a metal bar subject to
an external load. Failure of composite systems is of utmost importance in
naval, aeronautics and space industry (Reichhardt, 1996). By the term composite, we
refer to materials with heterogeneous microscopic structures and also to assemblages
of macroscopic elements forming a super-structure. Chemical and nuclear
plants suffer from cracking due to corrosion either of chemical or radioactive
origin, aided by thermal and/or mechanical stress. 

Despite the large amount of
experimental data and the considerable effort that has been undertaken by
material scientists (Liebowitz, 1984), many questions about fracture have
not been answered yet. There is no comprehensive understanding of rupture
phenomena but only a partial classification in
restricted and relatively simple situations. This lack of fundamental
understanding is indeed reflected in the absence of reliable prediction methods for
rupture, based on a suitable monitoring of the stressed system.
Not only is there a lack of non-empirical understanding of the reliability
of a system, but also the empirical laws themselves have often limited value.
The difficulties stem from the complex interplay between heterogeneities
and modes of damage and the
possible existence of a hierarchy of characteristic scales (static and dynamic).

Many material ruptures occur by a `one crack' mechanism and a lot of effort is 
being devoted to the understanding, detection and prevention of the nucleation of
the crack (Fineberg and Marder, 1999; Bouchaud, 2003).
Exceptions to the `one crack' rupture mechanism are heterogeneous 
materials such as fiber composites, 
rocks, concrete under compression, ice, tough ceramics and
materials with large distributed residual stresses. The common property 
shared by these systems is the existence of large inhomogeneities, 
that often limit the use of homogeneization or effective medium theories for the 
elastic and more generally the mechanical properties.
In these systems, failure may occur as the culmination of a progressive
damage involving complex interactions between multiple defects and growing micro-cracks.
In addition, other relaxation, creep, ductile, or plastic
behaviors, possibly coupled with corrosion effects may come into play.
Many important practical applications involve the coupling between 
mechanical and chemical effects with the competition between several characteristic
time scales. Application of stress may act as a catalyst of
chemical reactions (Gilman, 1996) or, reciprocally, chemical
reactions may lead to bond weakening 
(Westwood, Ahearn and Mills, 1981) and thus promote
failure. A dramatic example is the aging of present aircrafts due to repeating
loading in a corrosive environment (National Research Council, 1997).
The interaction between multiple defects and the existence of 
several characteristic scales present a considerable challenge to the modeling
and prediction of rupture. Those are the systems and problems on
which the interdisciplinary marriage with statistical physics has brought
new fruitful ideas that we now briefly present.

\begin{flushleft}
{\large\bf Creep rupture}
\vspace{0.05em}
\end{flushleft}

There are many different conditions under
which a material can rupture: 
constant strain rate, or stress, or stress rate, or more complex
strain/stress histories (involving also other control parameters
such as temperature, water content, chemical activity, and so on).
The situation in which a stress is imposed is very frequent
in mechanics (constant weight) and leads to the phenomenon of creep
(also known as `static fatigue').
A stress step leads in general to a strain response and other
observable changes such as acoustic emissions (see 
for a review (El Guerjouma et al., 2001)). Understanding 
damage and rupture of a material subjected to a constant stress is
thus a good starting point. For industrial applications, 
creep experiments are not always practical because they require 
adjusting the stress to sub-critical levels such that one does not 
wait too long before interesting processes (including eventually rupture)
are monitored. Accelerated tests, which yield information more
quickly, include step-stress and ramp-stress loading (Nelson, 1990).

As we said, time-dependent deformation of a material subjected to a constant stress
level is known as creep. In creep, the stress is below the mechanical
strength of the material, so that the rupture does not occur upon
application of the load. It is by waiting a sufficiently long time that
the cumulative strain may finally end in catastrophic rupture. Creep is
all the more important, the larger the applied stress and the higher the
temperature. The time to creep rupture is found in a large variety of
materials to be controlled by the stress sign and magnitude, temperature
and microstructure. 

Creep is often divided into three regimes: (i) the primary creep regime
corresponds to a decay of the strain rate following the application of the
constant stress, which can often be described by the so-called Andrade's law
(a power law decay with time); (ii) the secondary regime describes 
an (often very long) cross over, characterized by an approximately
constant strain rate, towards the (iii) tertiary creep 
regime in which the strain rate
accelerates up to rupture. Andrade's law for the strain rate is similar to the power-law
relaxation of the aftershock seismic activity triggered by the
stress change induced by a previous earthquake, known as Omori's law 
(Omori, 1894). In creep experiments, Omori's law describes the decay of the 
rate of acoustic emissions in the primary regime.
Creep experiments are thus interesting both because
they constitute standard mechanical tests of long-time properties of structures
and because of the power laws reminiscent of the critical behavior
of complex self-organizing systems that have become popular paradigms, as
discussed below.

Studies of the creep rupture phenomena have been performed
through direct experiments
(Agbossou, Cohen and Muller, 1995; Liu and Ross, 1996; Guarino et al., 2002;
Lockner, 1998) as well as through
different models (Miguel et al., 2002; Ciliberto, Guarino and Scorretti, 2001;
Kun et al., 2003; Hidalgo, Kun and Herrmann, 2002; Main, 2000;
Politi, Ciliberto and Scorretti, 2002; Pradhan and Chakrabarti, 2003;
Turcotte, Newman and Shcherbakov, 2003; Shcherbakov and Turcotte, 2003;
Saichev and Sornette, 2005). If a lot of works were devoted to
homogeneous materials like metals and ceramics, many recent studies
are concerned with heterogeneous materials like composites and rocks
(Agbossou, Cohen and Muller, 1995; Liu and Ross, 1996; Guarino et al., 2002;
Lockner, 1998). The knowledge of the
failure properties of composite materials are of great importance
because of the increasing number of applications for composites in
engineering structures. The long-term behavior of these materials,
especially polymer matrix composites is a critical issue for many modern
engineering applications such as aerospace, biomedical and civil
engineering infrastructure. The primary concerns in long-term
performance of composite materials  are in obtaining critical
engineering properties that extend over the projected lifetime of the
structure. Viscoelastic creep and creep-rupture behaviors are among the
critical properties needed to assess long-term performance of polymer-based 
composite materials. The knowledge of these critical properties is
also required to design material microstructures which can be used to
construct highly reliable components. For heterogeneous materials, the
underlying microscopic failure mechanism of creep rupture is very
complex depending on several characteristics of the specific types of
materials. Beyond the development of analytical and numerical models,
which predict the damage history in terms of the specific parameters of
the constituents, another approach is to study the similarity of creep
rupture with phase transitions phenomena as summarized here. This approach
tackles the large range of scales involved in the damage evolution 
by using coarse-grained models describing the mechanism of creep,
damage and precursory rupture by averaging over the microscopic degrees
of freedom to retain only a few major ingredients that are thought to be
the most relevant. By comparing the predictions of a hierarchy of models,
from simple to elaborate, it is then
possible to assess what are the relevant ingredients.

\begin{figure}
\includegraphics[width=0.35\textwidth,angle=-90]{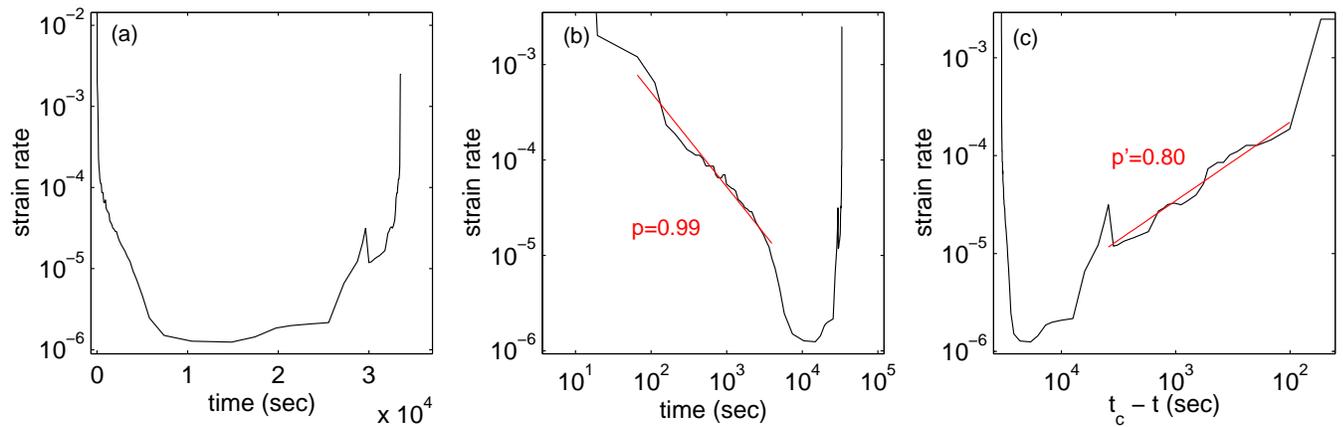}
\caption{\label{dedtSMC} Creep strain rate for a
Sheet Molding Compound (SMC) composite 
consisting in a combination of polyester resin, calcium carbonate filler, 
thermoplastic additive and random oriented short glass reinforced fibres. 
The creep experiment was performed at a stress of $48$ MPa and
a temperature $T=100^o$C, below the glass transition,
at the GEMPPM, INSA LYON, Villeurbanne, France.
The stress was increased progressively and reached a constant value after about 
17 sec. 
Left panel: full history in linear time scale; middle panel: time is shown in
logarithmic scale to test for the existence of a power law relaxation regime; 
right panel: time is shown in the logarithm of the time to rupture time $t_c$ such
that a time-to-failure power law (\ref{tc1}) is qualified as a straight line. 
Reproduced from (Nechad et al., 2004).}
\end{figure}

A recent experimental work on heterogenenous structural materials,
conducted in GEMPPM at INSA Lyon,  illustrates this approach (Nechad et al., 2005). 
Figure \ref{dedtSMC} shows a
rapid and continuous decrease of the strain rate $de/dt$
in the primary creep regime, which can be described by Andrade's law 
(Omori's law for the acoustic emissions)
\begin{equation}
{de\over dt} \sim {1 \over t^p}~,
\label{pc1}
\end{equation}
with an exponent $p$ smaller than or equal to one.
A quasi-constant strain rate (steady-state or secondary creep) is
observed over an important part 
of the total creep time and then followed by
an increasing creep rate (tertiary creep regime) culminating in fracture. 
Creep strains at fracture are large with values from a few percent up to 40\% for such
composite samples.
The acceleration of the strain rate before failure is well fitted by a
power-law singularity
\begin{equation}
{de\over dt} \sim {1 \over (t_c - t)^{p'}}~,
\label{tc1}
\end{equation}
with an exponent $p'$ smaller than or equal to one. 
The critical time $t_c$ determined from the fit of the data
with expression (\ref{tc1}) is generally close
to the observed failure time (within a few seconds). 
The same temporal evolution is generally obtained for the 
acoustic emission activity as for the strain rate.  
The same patterns are obtained when plotting the acoustic emission event 
rate or the rate of acoustic emission energy. There are 
much larger fluctuations for the energy
rate than for the event rate, due to the large distribution of 
acoustic emission energies, but the crossover time between primary creep and tertiary
creep, and the values of $p$ and $p'$ are similar for the 
acoustic emission event
rate and for the acoustic emission energy rate. This suggests that the amplitude 
distribution does not depend on time, a conclusion which is verified experimentally.
How can one rationalize all these observations?

\begin{flushleft}
{\large\bf The role of heterogeneities and disorder}
\vspace{0.05em}
\end{flushleft}

First, we need to define more precisely what is meant by `heterogeneity' or `disorder'.
Disorder may describe the existence of a distribution (say Weibull-like) of 
material strength, and/or of their elastic properties, as well as the presence of
internal surfaces such as fiber-matrix interfaces, 
voids and microcracks (or internal microdefects). In this sense, a kevlar-matrix
or carbon-matrix composite would behave more like a heterogeneous system than
a homogeneous matrix. 
There is not a unique way of defining the amplitude of disorder,
since the classification depends on how the mechanics and physics respond to the 
heterogeneity. It can in fact be shown from a theorem of 
Von Neumann and Morgenstern (1947) that
the existence of possible correlations in the disorder
prevents the existence of a unique absolute measure of disorder amplitude. In other words,
the measure of disorder is relative to the problem. 
In practice, it can usually be quantified by some measure of 
the contrast between material and strength properties of components of the systems, 
weighted by their relative concentrations and their scales. When 
disorder is uncorrelated in space,
a reasonable measure of its amplitude is the width or standard deviation
(when it exists) of its distribution. The correlation length of the disorder and
the characteristic sizes and their distribution are also
important variables as they control the length scales that are relevant for the
stress heterogeneity. A consequence is the size/volume effect,
which is a very important practical subject. 

As already mentioned, the key parameter controlling the nature
of damage and rupture is the degree and nature of disorder.
This was considered early by Mogi (1969), who showed experimentally on a variety
of materials that, the larger
the disorder, the stronger and more useful are the precursors to rupture. For a long time,
the Japanese research effort for earthquake prediction and risk assessment
was based on this very idea (Mogi, 1995). A quantification of this idea of 
the role of heterogeneities on the nature of rupture has been obtained with
a two-dimensional spring-block model 
with stress transfer over a limited range and with the
democratic fiber bundle model (Andersen, Sornette and Leung, 1997).
These models do not claim realism but attempt rather to capture the 
interplay of heterogeneity and of the stress transfer mechanism.
It was found that 
heterogeneity plays the role of a relevant field (in the language
of the statistical physics of critical phase transitions): systems with limited
stress amplification exhibit a tri-critical transition (Aharony, 1983), 
from a Griffith-type abrupt rupture (first-order) regime to a
progressive damage (critical) regime as the disorder
increases. In the two-dimensional
spring-block model of surface fracture (Andersen, Sornette and Leung, 1997), 
the stress can be released
by spring breaks {\it and} block slips.
This spring-block model may represent schematically the experimental
situation where a balloon covered with paint or dry resin is progressively
inflated. An industrial application may be for instance a metallic tank with carbon
or kevlar fibers impregnated in a resin matrix wrapped up around it  which is
slowly pressurized (Anifrani et al., 1995). As a consequence, it elastically deforms,
transferring tensile stress to the overlayer. Slipping (called fiber-matrix
delamination) and cracking can thus occur in the overlayer. 
In (Andersen, Sornette and Leung, 1997),
this process is modeled by an array of blocks which represents the overlayer on a coarse
grained scale in contact with a surface with solid friction contact. The solid
friction will limit stress amplification.
The fact that the disorder is so relevant as to create the
analog of a tri-critical behavior can be traced back to the existence of solid
friction on the blocks which ensures that the elastic forces in the springs are
carried over a bounded distance (equal to the size of a slipping `avalanche')
during the stress transfer induced by block motions. 
There are similarities between this model and models of quasi-periodic matrix
cracking in fibrous composites and of fragmentation of fibers in the so-called
`single-filament-composite' test. This last model has been extensively 
developed and extended in a global and local load-sharing framework
(Curtin, 1991; 1998; Ibnabdeljalil and Curtin, 1998).

In the presence of long-range elasticity, disorder is found to be always
relevant leading to a critical rupture. However, 
the disorder controls the width of the critical region (Sornette and Andersen, 1998). The
smaller it is, the smaller will be the critical region, which
may become too small to play any role in practice. This has been confirmed by
simulations of the `thermal fuse model' described below
(Sornette and Vanneste, 1992): the damage rate 
on the approach to failure for different disorder can be rescaled 
onto a universal master curve.

\begin{flushleft}
{\large\bf Qualitative physical scenario: from diffuse damage to global failure}
\vspace{0.05em}
\end{flushleft}

The following qualitative physical picture for the progressive damage of an heterogeneous
system leading  to global failure has emerged from a large variety of 
theoretical, numerical and experimental works
(see for instance (Lei et al., 1999; 2000; Tang et al., 2000)). 
First, single isolated defects and
microcracks nucleate which then, with the increase of load or time of loading, both grow
and multiply leading to an increase of the density of defects per unit volume.
As a consequence, defects begin to merge until a `critical density' is
reached. Uncorrelated percolation (Stauffer and Aharony, 1992) 
provides a starting modeling
point valid in the limit of very large disorder (Roux et al., 1988). For
realistic systems, long-range correlations transported by the stress field
around defects and cracks make the problem much more subtle. Time dependence is
expected to be a crucial aspect in the process of correlation building in these
processes. As the damage increases, a new `phase' appears, where micro-cracks 
begin to merge leading to screening and other cooperative effects. 
Finally, the main fracture is formed causing global failure. The nature of
this global failure may be abrupt (`first-order') or `critical' 
depending of the strength of heterogeneity as well as load transfer and 
stress relaxation mechanisms. In the `critical' case,
the failure of composite systems
may often be viewed, in simple intuitive terms, as the result of a `correlated 
percolation process.' However, the challenge is to describe the transition from
damage and corrosion processes at a microscopic level to macroscopic
failure.

\begin{flushleft}
{\large\bf Scaling and critical point}
\vspace{0.05em}
\end{flushleft}

Motivated by the multi-scale nature of ruptures in heterogeneous systems and by
analogies with the percolation model (Stauffer and Aharony, 1992), 
statistical physicists suggested in the mid-1980s
that rupture of sufficiently heterogeneous media would exhibit some
universal properties, in a way maybe similar to critical phase transitions
(de Arcangelis, Redner and Herrmann, 1985;
Duxbury, Beale and Leath, 1986; Gilabert et al., 1987). The idea
was to build on the knowledge accumulated in statistical physics on
the so-called $N-$body problem and cooperative effects in order to describe
multiple interactions between defects. However, 
most of the models were drastically simplified and essentially all of them quasi-static
with rather unrealistic loading rules (Herrmann and Roux, 1990; Meakin, 1991).
Suggestive scaling laws, including multifractality,
were found to describe size effects and damage properties
(Herrmann and Roux, 1990; Hansen, Hinrichsen and Roux, 1991), 
but the relevance to real materials was not convincingly
demonstrated with a few exceptions (e.g., percolation theory to explain
the experimentally based  Coffin-Manson law of low cycle fatigue 
(Br\'echet, Magnin and Sornette, 1992) or the Portevin-Le Chatelier
effect in diluted alloys (Bharathi and Ananthakrishna, 2002)).

With numerical simulations and 
perturbation expansions, 
Hansen, Hinrichsen and Roux (1991) (see also Herrmann and Roux, 1990) have
used this class of quasi-static rupture models (with short-range as
well as long-range interactions) to classify three possible
rupture regimes, as a function of the concentrations of weak versus strong elements
in the system. Specifically, the distribution $p(x)$ of rupture thresholds $x$
of elements of the discretized systems was parameterized as follows:
$p(x) \sim x^{\phi_0 -1}$ for $x \rightarrow 0$ and 
$p(x) \sim x^{-(1+\phi_{\infty})}$ for $x \rightarrow +\infty$.
Then, the three regimes
depend on the relative value of the exponents $\phi_0$ and $\phi_{\infty}$ compared with 
two critical values $\phi_0^c$ and $\phi_{\infty}^c$. The `weak disorder' regime
occurs for $\phi_0 >\phi_0^c$ (few weak elements) and $\phi_{\infty}>\phi_{\infty}^c$ 
(few strong elements) and boils down
essentially to the nucleation of a `one-crack' run-away. For 
$\phi_0 \leq \phi_0^c$ (many weak elements) and $\phi_{\infty}>\phi_{\infty}^c$
(few strong elements), the rupture is 
controlled by the weak elements, with important size effects.
The damage is diffuse but presents a structuration at large scales. 
For $\phi_0 > \phi_0^c$ (few weak elements) and $\phi_{\infty} \leq \phi_{\infty}^c$
(many strong elements), the
rupture is controlled by the strong elements\,: the final
damage is diffuse and the density of broken elements goes to a non-vanishing
constant. This third case is very similar to the percolation models of rupture:
Roux et al. (1988) have indeed shown that percolation is retrieved
in the limit of very large disorder.

Beyond quasi-static models, the `thermal fuse model' of
Sornette and Vanneste (1992) was
the first one with a realistic dynamical evolution law
for the damage field. 
It was initially formulated in the framework
of electric breakdown: when subjected to a given current, all fuses
in a network heat up due to a
generalized Joule effect (with exponent $b$); in the presence of 
heterogeneity in the conductances of the fuses, one of them will  
eventually breaks down first when its temperature reaches
the melting threshold. Its current is then 
immediately distributed in the remaining fuses according to Kirchoff law.
The model was later reformulated by showing that it
is exactly equivalent to a (scalar) antiplane mechanical model of rupture with 
elastic interaction in which the temperature becomes a local damage variable
(Sornette and Vanneste, 1994).
This model accounts for space-dependent elastic and rupture properties, has a realistic
loading (constant stress applied at the beginning of the simulation, for instance) 
and produces growing interacting micro-cracks with an organization 
which is a function of the damage-stress law. 
This model is thus a statistical generalization with quenched disorder of
homogeneization theories of damage (Chaboche, 1995; Maire and Chaboche, 1997).
In a creep experiment 
(constant applied stress), the total rate of damage in the late stage
of evolution, as measured for instance by
the elastic energy released per unit time $dE/dt$, is found on average to 
increase as a power law similar to expression (\ref{tc1}),
\begin{equation}
dE/dt \sim 1/(t_c-t)^{\alpha}~,
\label{eq1}
\end{equation}
of the time-to-failure $t_c-t$ in the later stage. This behavior reproduces
the tertiary creep regime culminating in the global rupture at $t_c$.
In this model, rupture is found 
to occur as the culmination of the progressive nucleation, growth and fusion
between microcracks, leading to a fractal network. Interestingly, the
critical exponents (such as $\alpha>0$)
are non-universal and vary as a function of the damage law (exponent $b$). This
model has since then been found to describe correctly the experiments on 
the electric breakdown of insulator-conducting composites 
(Lamaign\`ere, Carmona and Sornette, 1996).
Another application of the thermal fuse model is the damage
by electromigration of polycristalline metal films (Bradley and Wu, 1994).
See also (Sornette and Vanneste, 1994) 
for relations with dendrites and fronts propagation.

The concept that rupture in heterogenous materials is a genuine
critical point, in the sense of phase transitions in statistical
physics, was first articulated by Anifrani et al. (1995), based
on experiments on industrial composite structures. In this framework,
the power law (\ref{eq1}) is interpreted as analogous to a diverging
susceptibility in critical phase transitions.
It was found that the critical behavior may correspond to an acceleration of the 
rate of energy release or to a deceleration, depending on the nature and range of the 
stress transfer mechanism and on the loading procedure. 
Symmetry arguments as well as the concept of
a hierarchical cascade of damage events led 
in addition to suggest theoretically and verify
experimentally that the power law behavior (\ref{eq1}) of the time-to-failure
analysis should be corrected for the presence of log-periodic modulations 
(Anifrani et al., 1995). This `log-periodicity' can be
shown to be the signature 
of a hierarchy of characteristic scales in the
rupture process. This hierarchy can be generated dynamically by a
cascade of sub-harmonic bifurcations (Huang et al., 1997).
These log-periodic corrections to scaling 
amount mathematically to taking the critical exponent $\alpha=\alpha'+i \alpha''$ complex,
where $i^2=-1$ (Sornette, 1998). This has led to the development of a 
powerful predictive scheme ((Le Floc'h and Sornette, 2003) and see below).
The critical rupture concept can be seen as a non-trivial generalization
of the dimension analysis based on Buckingham theorem
and the asymptotic matching method proposed by Bazant (1997) to model
size effect in complex materials,
in the same way that Barenblatt (1987)'s second-order similitude generalizes
the naive similitude of first-order (or simple analytical behavior) of 
standard dimensional analysis, or in the same way the non-analytical
behavior characterizing critical phase transitions generalizes the mean-field
behavior of Landau-Ginzburg theory. Acharyya and Chakrabarti (1996) have 
shown how to define a ``breakdown
susceptibility'' during the progressive damage of model systems when
subjected to local short-duration impulses and how the breakdown point
can then be located in advance by extrapolating this breakdown susceptibility. 

Numerical simulations on two-dimensional heterogeneous
systems of elastic-brittle elements 
have confirmed that, near the global failure point, 
the cumulative elastic energy released during fracturing of heterogeneous
solids with long-range elastic interactions
follows a power law with log-periodic corrections to the leading term
(Sahimi and Arbati, 1996).
The presence of log-periodic correction to scaling in the elastic energy released
has also been demonstrated numerically for the thermal fuse model 
(Johansen and Sornette, 1998) using a novel
averaging procedure, called the `canonical ensemble averaging.' This averaging
technique accounts for the dependence of the critical rupture time $t_c$ 
on the specific disorder realization of each sample.
A recent experimental study of rupture of fiber-glass
composites has also confirmed the critical scenario (Garcimartin et al., 1997). 
A systematic analysis of industrial pressure tanks brought to rupture
has also confirmed the critical rupture concept and the presence
of significant log-periodic structures, that are useful for prediction
(Johansen and Sornette, 2000). Through a series of computer and
laboratory simulations and table-top experiments, Chakrabarti and
Benguigui (1997) have presented a useful synthesis of basic modeling principles
borrowing from statistical physics
putting in perspective three case studies: electrical failures like fuse
and dielectric breakdown, mechanical fractures, and earthquakes. 
Their work also emphasizes the critical rupture concept 
(Acharyya and Chakrabarti,1996; 
Banerjee and Chakrabarti, 2001; Pradhan and Chakrabarti, 2002). 

Let us also mention the work of Ramanathan and Fisher (1998): using
analytical calculations and by numerical simulations, they compare
the nature of the onset of a single crack motion in an heterogeneous
material when neglecting or taking into account the dynamical wave
stress transfer mechanism. In the quasistatic limit with instantaneous stress transfer,
the crack front is found to undergo a dynamic critical phenomenon, with a
second-order-like transition from a pinned to a moving phase as the
applied load is increased through a critical value. Real elastic waves
lead to overshoots in the stresses above their eventual static
value when one part of the crack front moves forward. Simplified models
of these stress overshoots showed an apparent jump in the velocity of 
the crack front directly to a
nonzero value. In finite systems, the velocity also shows hysteretic
behavior as a function of the loading. These results suggest a
first-order-like transition (Ramanathan and Fisher, 1998).

\begin{flushleft}
{\large\bf Creep rupture: models}
\vspace{0.05em}
\end{flushleft}

Let us come back to the experiments shown in Figure \ref{dedtSMC}.
There are many models, at the interface between standard mechanical 
approaches and statistical physics, which attempt to capture these observations.
Vujosevic and Krajcinovic (1997), Turcotte, Newman and Shcherbakov (2003),
Shcherbakov and Turcotte (2003) and Pradhan and Chakrabarti (2004)
used systems of elements or fibers
within a probabilistic framework (corresponding to so-called annealed or
thermal disorder) with a hazard rate function
controlling the probability of rupture for a given fiber as
a function of the stress applied to that fiber. 
Turcotte, Newman and Shcherbakov (2003) obtained a finite-time
singularity of the strain rate before failure in fiber bundle models
by postulating a power law dependence of the hazard rate
controlling the probability of rupture for a given fiber as
a function of the stress applied to that fiber.
Shcherbakov and Turcotte (2003) studied the same model
and recovered a power-law singularity of the strain rate 
for systems subjected to constant or increasing stresses with an
exponent $p'=4/3$ larger than the experimental results. 
Using energy conservation and the requirement of non-negative entropy
change, Lyakhovsky, Ben-zion and Agnon (1997) derived
an evolution equation for the density of microcracks similar
to that of Turcotte, Newman and Shcherbakov (2003) for a fiber bundle model.
Ben-Zion and Lyakhovsky (2002) derived analytically the existence of power laws
describing the time-dependent increase of the singular strain and
the accelerated energy release in the tertiary regime using the continuum-based damage
approach of Lyakhovsky, Ben-zion and Agnon (1997). Sammis and Sornette (2002)
give an exhaustive review of the mechanisms giving rise to the power law
tertiary regime, with application to earthquakes.
Vujosevic and Krajcinovic (1997) also found a 
power-law acceleration in two-dimensional
simulations of elements and in a mean-field democratic load sharing model, 
using a stochastic hazard rate, but they do not obtain Andrade's law
in the primary creep regime. 
Shcherbakov and Turcotte (2003) were able to obtain Andrade law
only in the situation of a system subjected to a constant applied
strain (stable regime). But then, they did not have a global rupture
and they did not obtain the critical power law preceding rupture.
Thus, the models described above do not reproduce at the same
time Andrade's law for the primary regime and a power-law singularity
before failure.
Miguel et al. (2002) reproduced Andrade's law with $p\approx 2/3$ 
in a numerical model of interacting dislocations, but their model
does not reproduce the tertiary creep regime (no global failure).

Several creep models consider the democratic fiber bundle model (DFBM) 
with thermally activated failures of fibers. 
Pradhan and Chakrabarti (2004) considered the DFBM and added a
probability of failure per unit time for each fiber which depends 
on the amplitude of a thermal noise and on the applied stress.
They computed the failure time as a function of the applied stress and noise 
level but they did not discuss the temporal evolution of the strain rate.
Ciliberto, Guarino and Scorretti (2001) and 
Politi, Ciliberto and Scorretti (2002)
considered the DFBM in which a random fluctuating force is added on each fiber
to mimic the effect of thermal fluctuations.  Ciliberto, Guarino and Scorretti (2001)
showed that this simple
model predicts a characteristic rupture time given by an Arrhenius law
with an effective temperature renormalized (amplified) by the quenched
disorder in the distribution of rupture thresholds. Saichev and Sornette (2005) 
showed that this model predicts Andrade's law as well as a power 
law time-to-failure for the rate of fiber rupture with $p=p'=1$, with
logarithm corrections (which may give apparent exponents $p$ and $p'$ smaller than $1$). 

A few other models reproduce both a power-law relaxation in the
primary creep and a finite time singularity in the tertiary regime.
Main (2000) reproduced a power-law relaxation (Andrade's law)
followed by a power-law singularity of the strain rate before failure by
superposing two processes of subcritical crack growth, with different 
parameters. A first mechanism with negative feedback dominates in the primary
creep and the other mechanism with positive feedback gives the
power-law singularity close to failure. Lockner (1998) gave
an empirical expression for the strain rate as a function of the
applied stress in rocks, which reproduces, among other properties, Andrade's law
with $p=1$ in the primary regime and a finite-time singularity leading
to rupture.

Kun et al. (2003) and Hidalgo, Kun and Herrmann (2002) studied numerically 
and analytically a model of visco-elastic fibers, with deterministic dynamics and
 quenched disorder. They considered different ranges of interaction between fibers
(local or democratic load sharing). Kun et al. (2003) derived
the condition for global failure in the system and the evolution of the failure
time as a function of the applied stress in the unstable regime, and analysed
the statistics of inter-event times in numerical simulations of the model.
Hidalgo, Kun and Herrmann (2002) derived analytically the expression for the strain 
rate as a function of time.
This model reproduces a power-law singularity of the strain rate before failure 
with $p'=1/2$ in the case of a uniform distribution of strengths,
but is not able to explain Andrade's law for the primary creep.
This model gives a power-law decay of the strain rate in the primary creep regime 
only if the stress is  at the critical point, but with an exponent $p=1/2$ smaller
than the experimental values.
Nechad et al. (2005) developed a variant of this model in which
a composite system is viewed as made of a large set of representative elements (RE),
each representative element comprising many fibers with their interstitial matrix.
Each RE is endowed with a visco-elasto-plastic rheology with 
parameters which may be different from one element to another. The parameters 
characterizing each RE are frozen and do not evolve with time (so-called
quenched disorder). Specifically, each RE is modeled as an Eyring dashpot in
parallel with a linear spring. The Eyring rheology is standard for fiber
composites (Liu and Ross, 1996). It consists, at the microscopic level,
in adapting to the matrix rheology the theory of reaction rates
describing processes activated by crossing potential barriers.
With these sole ingredients, the model recovers the three primary, secondary
and tertiary regimes with exponents $p=1$ (defined in expression (\ref{pc1}))
and $p'=1$ (defined in expression (\ref{tc1})).
These solutions for the primary and tertiary regimes 
are basically of the same form with $p=p'=1$
as the Langevin-type model solved by Saichev and Sornette (2005); this may not be surprising
since the Eyring rheology describes, at the microscopic level,
processes activated by crossing potential barriers, which are explicitely
accounted for in the thermal fluctuation force model (Saichev and Sornette, 2005).
The key ingredients leading to these results are the broad (power law)
distribution of rupture thresholds and the nonlinear Eyring rheology in a
Kelvin element. Nechad et al.'s model is a macroscopic
deterministic effective description of the experiments. In contrast, the modeling 
strategy of Ciliberto, Guarino and Scorretti (2001), of 
Politi, Ciliberto and Scorretti (2002) and of Saichev and Sornette (2005)
emphasizes the interplay between 
microscopic thermal fluctuations and frozen heterogeneity. Qualitatively,
Nechad et al.'s model is similar to a deterministic macroscopic Fokker-Planck description 
while the thermal models of Ciliberto, Guarino and Scorretti (2001), of 
Politi, Ciliberto and Scorretti (2002) and of Saichev and Sornette (2005) are reminiscent 
of stochastic Langevin models. It is well-known in statistical physics that 
Fokker-Planck equations and Langevin equations are exactly equivalent 
for systems at equilibrium and just
constitute two different descriptions of the same processes, and their
correspondence is associated with the general fluctuation-dissipation
theorem. Similarly, the
encompassing of both the Andrade relaxation law in the primary creep regime and of
the time-to-failure power law singularity in the tertiary regime by
Nechad et al.'s model and by the thermal model solved in 
(Saichev and Sornette, 2005) suggests a deep connection 
between these two levels of description for creep and damage processes.

\begin{flushleft}
{\large\bf Toward rupture prediction}
\vspace{0.05em}
\end{flushleft}

\begin{figure}
\includegraphics[width=\textwidth]{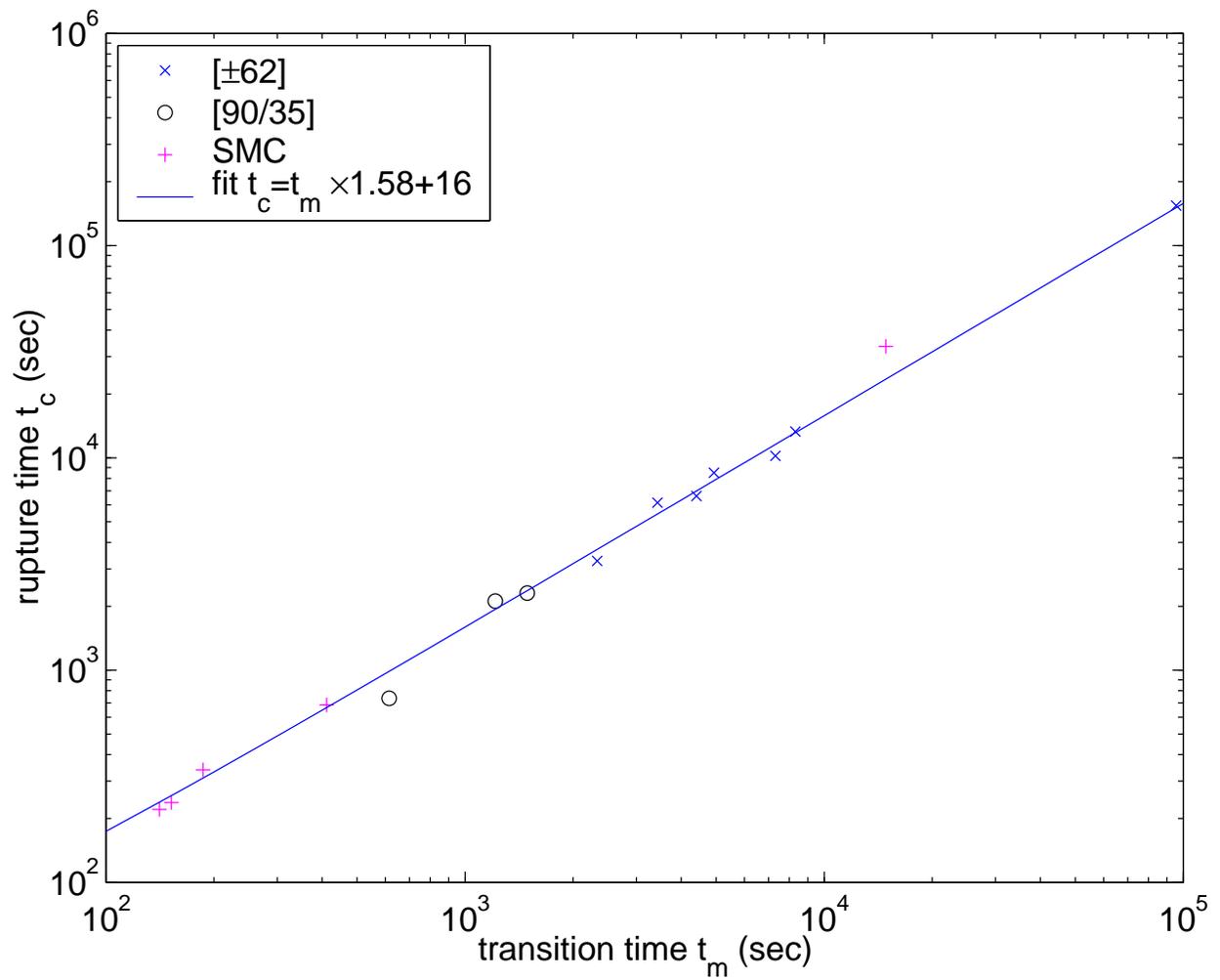}
\caption{\label{tmtc} Relation between the time $t_m$ of the minima of  the 
strain rate and the rupture time $t_c$, for all samples investigated
in (Nechad et al., 2004).} 
\end{figure}

There is a huge variability of the failure time from one sample to another one, 
for the same applied stress, as shown in
Figure \ref{tmtc}. This implies that one cannot 
predict the time to failure of a sample using an empirical relation between the 
applied stress and the time of failure. There is however another approach suggested 
by Figure \ref{tmtc} as proposed by Nechad et al. (2005). 
It shows the correlation between the transition 
time $t_m$ (minimum of the strain rate) and the rupture time $t_c \approx t_{end}$ 
and shows that $t_m$ is about $2/3$ of the rupture time $t_c$. 
This suggests a way to predict the failure time from the observation of the strain 
rate during the primary and secondary creep regimes, before the acceleration of 
the damage during the tertiary creep regime leading to the rupture of the sample.
As soon as a clear minimum is observed, the value of $t_m$ can be measured and that 
of $t_c$ deduced from the relationship shown in Figure \ref{tmtc}. 
However, there are some cases where the minima is not well defined, 
for which the first (smoothed) minimum is
followed by a second similar one. In this case, the application of the relationship
shown in Figure \ref{tmtc} would lead to a pessimistic prediction for the lifetime
of the composite.

The observation that the failure time is correlated with the $p$-value 
and the duration of the 
primary creep suggests that, either a single mechanism is responsible both 
for the decrease of the strain rate during primary creep and for the 
acceleration of the damage during the tertiary creep or, if the mechanisms 
are different nevertheless, the damage that occurs in the primary regime 
impacts on its subsequent evolution in the secondary and tertiary regime, 
and therefore on $t_c$.
In contrast, using a fit of the 
acoustic emission activity by a power-law to 
estimate $t_c$ according to formula (\ref{eq1}) 
works only in the tertiary regime and thus does not exploit the information
contained in the deformation and in the acoustic emissions of the primary and secondary 
regimes which cover 2/3 to 3/4 of the whole history. In practice, one needs
at least one order of magnitude in the time $t_c-t$ to 
estimate accurately $t_c$ and $p'$,
which means that, if the power-law acceleration regime starts immediately 
when the stress is applied (no primary creep), one cannot predict the rupture time
using a fit of the damage rate by equation (\ref{eq1}) before 90\% of the failure time.
If, as observed in the experiments of Nechad et al. (2005), 
the tertiary creep regime starts only at about 
63\% of $t_c$, then one cannot predict the rupture time using a fit of the damage 
rate before 96\% of the failure time. This limitation was the motivations
for the development of formulas that interpolate between the primary
and tertiary regimes beyond the pure power law (\ref{eq1}) using log-periodic
corrections to scaling (Anifrani et al., 1995; Johansen and Sornette, 2000; 
Gluzman et al., 2001; Moura and Yukalov, 2002; Gauthier et al., 2002;
Yukalov et al., 2004).

In particular, Anifrani et al. (1995) have introduced a method
based on log-periodic correction to the critical power law
which has been used extensively by 
the European Aerospace company A\'erospatiale (now EADS) 
on pressure tanks made of kevlar-matrix
and carbon-matrix composites embarked on the European Ariane 4 and 5 rockets. 
In a nutshell, the method consists in this application in 
recording acoustic emissions under constant stress rate
and the acoustic emission energy as a function of
stress is fitted by the above log-periodic critical theory. One of the parameters is the
time of failure and the fit thus provides a `prediction' when the sample
is not brought to failure in the first test (Gauthier et al., 2002). 
The results indicate that a precision of  
a few percent in the determination of the stress at rupture 
is typically obtained using acoustic emission recorded about $20~\%$ below
the stress at rupture.
This has warranted the selection of this non-destructive
evaluation technique as the routine qualifying procedure in the industrial fabrication
process. This methodology and these experimental results have
been guided by the theoretical research over the years using the
critical rupture concept discussed above.
In particular, there is now a better understanding of the conditions, 
the mathematical properties and physical mechanisms at the basis of
log-periodic structures (Huang et al., 1997; Sornette, 1998; 2002; Ide and Sornette, 2002;
Zhou and Sornette, 2002). Another noteworthy approach already mentioned above for 
the prediction of rupture, which is inspired by statistical
physics, is the ``breakdown susceptibility'' introduced by 
Acharyya and Chakrabarti (1996). It requires monitoring the response
of the system when subjected to local short-duration impulses whose
nature depends upon the problem (stress, strain, temperature, electromagnetic and so on).

In summary, starting with the initial flurry of interest from the statistical physics
community on problems of material rupture, a new awareness of the many-body nature
of the rupture problem has blossomed. There is now a growing understanding
in both communities of the need for an interdisciplinary approach, improving
on the reductionist approach of both fields to tackle at the same time the
difficult modelling of specific properties of the microscopic structures and their interactions
leading to collective effects. Independently of the types of materials for
given applications, this approach will be crucial in making progress on 
the optimisation of the lifetime of materials (``durability'') and on the 
determination of the remaining life time of materials in use (``remaining potential'').


\begin{flushleft}
{\large\bf Bibliography}
\vspace{0.05em}
\end{flushleft}

Acharyya, M. and Chakrabarti, B.K., 1996.
Response of random dielectric composites and earthquake models to pulses --
prediction possibilities, {\it Physica A} 224, 254-266.

Acharyya, M. and Chakrabarti, B.K.,1996. Growth of breakdown susceptibility
in random composites and the stick-slip model of earthquakes -- Prediction
of dielectric breakdown and other catastrophes, {\it Phys. Rev. A} 53, 140-147;
Correction, {\it Phys. Rev. A} 54, 2174-2175.

Agbossou, A., I. Cohen and D. Muller, 1995.
Effects of interphase and impact strain rates on tensile off-axis behaviour 
of unidirectional glass fibre composite: experimental results, 
{\it Engineering Fracture Mechanics}, 52 (5), 923-935.

Aharony, A., 1983. Tricritical phenomena. {\it Lecture Notes in Physics}, 186, 209 (1983).

Andersen, J.V., D. Sornette and K.-T. Leung, 1997.
Tri-critical behavior in rupture induced by disorder.
{\it Phys. Rev. Lett.}, 78, 2140-2143.

Anifrani, J.-C., C. Le Floc'h, D. Sornette and B. Souillard, 1995.
Universal Log-periodic correction to renormalization group scaling for rupture stress
prediction from acoustic emissions. {\it J.Phys.I France}, 5, 631-638.

Banerjee, R. and Chakrabarti, B.K., 2001. Critical fatigue behaviour in brittle
glasses, {\it Bulletin of Materials Science} 24(2), 161-164.

Barenblatt, G.I., 1987. Dimensional Analysis (Gordon and Breach, New York).

Bazant, Z.P., 1997. Scaling of quasibrittle fracture: Asymptotic analysis.
{\it Int. J. Fracture}, 83, 19-40.

Bazant, Z.P., 1997. Scaling of quasibrittle fracture: 
Hypotheses of invasive and lacunar fractality, their critique and Weibull connection.
{\it Int. J. Fracture}, 83, 41-65. 

Ben-Zion, Y. and V, Lyakhovsky, 2002. Accelerated seismic 
release and related aspects of seismicity patterns on 
earthquake faults. {\it Pure \& Applied Geophysics}, 159(10), 2385-2412. 

Bharathi, M.S. and Ananthakrishna, G., 2002. Chaotic and 
power law states in the Portevin-Le Chatelier effect. 
{\it Europhysics Letters} 60, 234-240; 
Correction, 2003. Ibid 61, 430. 

Bouchaud, E., 2003. The morphology of fracture surfaces: 
A tool for understanding crack propagation in complex materials. 
{\it Surface Review \& Letters}, 10, 797-814. 

Bradley, R.M. and K. Wu, 1994. 
Dynamic fuse model for electromigration failure of polycrystalline metal films, 
{\it Phys. Rev. E}, 50, R631-R634. 

Br\'echet, Y., T. Magnin and D. Sornette, 1992.
The Coffin-Manson law as a consequence of the statistical nature of the
LCF surface damage, {\it Acta Metallurgica}, 40, 2281-2287.

Chaboche, J.L., 1995. A continuum damage theory with anisotropic and unilateral damage. 
{\it Recherche Aerospatiale}, 2, 139.

Chakrabarti, B.K. and Benguigui, L.G., 1997. 
Statistical physics of fracture and breakdown in
disordered Systems (Clarendon Press, Oxford).

Ciliberto, S., A. Guarino and R. Scorretti, 2001.
The effect of disorder on the fracture nucleation process.
{\it Physica D}, 158, 83-104.

Curtin, W.A., 1991. Exact theory of fibre fragmentation in a single-filament
composite, {\it J. Mater. Science}, 26, 5239-5253.

Curtin, W.A., 1998.
Size scaling of strength in heterogeneous materials, 
{\it Phys. Rev. Lett.}, 80, 1445-1448.

de Arcangelis, L., Redner, S., Herrmann, H.J., 1985.
A random fuse model for breaking processes,
{J. Physique Lettres}, 46, L585-590.

Duxbury, P.M., Beale, P.D., Leath, P.L., 1986.
Size effects of electrical breakdown in quenched random media,
{\it Phys. Rev. Lett.} 57, 1052-1055.

El Guerjouma, R., Baboux, J.C., Ducret, D., Godin, N., Guy, P., Huguet, S.,
Jayet, Y. and Monnier, T., 2001. Non-destructive evaluation of damage and failure
of fiber reinforced polymer composites using ultrasonic waves and
acoustic emission, {\it Advanced Engineering Materials}, 8, 601-608.

Fineberg J. and Marder, M., 1999. Instability in dynamic fracture.
{\it Physics Reports}, 313, 2-108. 

Garcimartin, A.,  Guarino, A.,  Bellon, L. and Ciliberto, S., 1997.
Statistical properties of fracture precursors.
{\it Phys. Rev. Lett.}, 79, 3202-3205.

Gauthier, J., C. Le Floc'h and D. Sornette, 2002.
Predictability of catastrophic events;  a new approach for Structural Health Monitoring
Predictive acoustic emission application on helium high pressure tanks. 
Proceedings of the first European workshop Structural Health Monitoring,
edited by D. Balageas (ONERA), pp. 926-930.
(http://arXiv.org/abs/cond-mat/0210418)

Gilabert, A., Vanneste C., Sornette D. and Guyon E., 1987.
The random fuse network as a model of rupture in a disordered medium,
{\it J. Phys. France}, 48, 763-770.

Gilman, J.J., 1996. Mechanochemistry. {\it Science}, 274, 65-65.

Gluzman, S., J.V. Andersen and D. Sornette, 2001.
Functional Renormalization Prediction of Rupture.
{\it Computational Seismology}, 32, 122-137.

Guarino, A., Ciliberto, S., Garcimartin, A. Zei, M., Scorretti, R.,
2002. Failure time and critical behaviour of fracture precursors in
heterogeneous materials. {\it Eur. Phys. J. B.} 26 (2), 141-151.

Hansen, A., E. Hinrichsen and S. Roux, 1991. Scale-invariant
disorder in fracture and related breakdown phenomena. {\it Phys. Rev. B}, 43, 665-678.

Herrmann, H.J. and S. Roux, (eds.), 1990.
{\it Statistical models for the fracture of disordered media},
(Elsevier, Amsterdam).

Hidalgo, R.C., Kun, F., Herrmann, H.J., 2002. 
Creep rupture of viscoelastic fiber bundles. {\it Phys.
Rev. E.}, 65 (3), 032502/1-4.

Huang, Y., G. Ouillon, H. Saleur and D. Sornette, 1997.
Spontaneous generation of discrete scale invariance in growth models.
{\it Phys. Rev. E}, 55, 6433-6447.

Ibnabdeljalil. M. and W.A. Curtin, 1997.
Strength and reliability of fiber-reinforced composites: Localized
load-sharing and associated size effects,
{\it Int. J. Sol. Struct.}, 34, 2649-2668.

Ide, K. and D. Sornette, 2002.
Oscillatory Finite-Time Singularities in Finance, Population and Rupture.
{\it Physica A}, 307 (1-2), 63-106.

Johansen, A. and D. Sornette, 1998. Evidence of discrete 
scale invariance by canonical averaging.
{\it Int. J. Mod. Phys. C}, 9, 433-447.

Johansen, A. and D. Sornette, 2000.
Critical ruptures. {\it Eur. Phys. J. B}, 18, 163-181.

Kun, F., Moreno, Y., Hidalgo, R.C., Herrmann, H.J., 2003. 
Creep rupture has two universality
classes. {\it Europhysics Letters}, 63 (3), 347-353.

Lamaign\`ere, L., F. Carmona and D. Sornette, 1996.
Experimental realization of critical thermal fuse rupture, 
{\it Phys. Rev. Lett.}, 77, 2738-2741.

Le Floc'h, C. and D. Sornette, 2003.
Predictive acoustic emission: Application on helium high pressure tanks,
Pr\'ediction des \'ev\`enements catastrophiques: une 
nouvelle approche pour le controle de sant\'e structurale,
Instrumentation Mesure Metrologie (published by Hermes Science)
RS serie I2M volume 3 (1-2), 89-97 (in french).

Lei, X.-L., K. Kusunose, O. Nishizawa, A. Cho and T. Satoh, 2000.
On the spatio-temporal distribution of acoustic emissions in two
granitic rocks under triaxial compression: the role of
pre-existing cracks, {\it Geophys. Res. Lett.}, 27, 1997-2000.

Lei, X.-L., K. Kusunose, M.V.M.S. Rao, O. Nishizawa and T. Satoh, 1999.
Quasi-static fault growth and cracking in homogenous brittle rock 
under triaxial compression using acoustic emission monitoring. 
{\it J. Geophys. Res.}, 105, 6127-6139.

Liebowitz, H., (ed.), 1984. {\it Fracture}.
New York, Academic Press. Vols. I-VII.

Liu, J.Y., Ross, R.J., 1996. Energy criterion for fatigue strength of
wood structural members. {\it Journal of Engineering Materials and
Technology}, 118 (3), 375-378.

Lockner, D. A., 1998. A generalized law for brittle deformation of
Westerly granite. {\it J. Geophys. Res.}, 103 (B3), 5107-5123.

Lyakhovsky, V., Y. Benzion and A. Agnon, 1997. Distributed damage, faulting
and friction. {\it J. Geophys. Res. (Solid Earth)},
102(B12), 27635-27649. 

Main, I. G., 2000. A damage mechanics model for power-law 
creep and earthquake aftershock and
foreshock sequences. {\it Geophys. J. Int.}, 142 (1), 151-161.

Maire, J.F. and J.L. Chaboche, 1997.
A new formulation of continuum damage mechanics (CDM) for composite materials.
{\it Aerospace Science and Technology}, 1, 247-257.

Meakin, P., 1991. Models for Material Failure and Deformation.
{\it Science}, 252 (5003), 226-234.

Miguel, M.C., Vespignani, A. Zaiser, M. Zapperi, S., 2002. 
Dislocation jamming and Andrade
creep. {\it Phys. Rev. Lett.}, 89 (16), 165501.

Mogi, K., 1969. Some features of recent seismic activity in and near Japan: 
activity before and after great earthquakes. {\it Bull. Eq. Res. Inst. Tokyo
Univ.}, 47, 395-417.

Mogi, K., 1995. Earthquake prediction research in Japan.
{\it J. Phys. Earth}, 43, 533-561.

Moura, A., Yukalov, V. I., 2002. Self-similar extrapolation 
for the law of acoustic emission before
failure of heterogeneous materials, Int. J. Fract. 118 (3), 63-68.

National Research Council, 1997.
Aging of U.S. Air Force Aircraft, Final Report from
the Committee on Aging of U.S. Air Force Aircraft, National Materials Advisory Board
Commission on Engineering and Technical Systems,
Publication NMAB-488-2, National Academy Press, Washington, D.C.

Nechad, H., A. Helmstetter, R. El Guerjouma and D. Sornette, 2005.
Andrade and Critical Time-to-Failure Laws in Fibre-Matrix Composites:
Experiments and Model, in press in  
Journal of Mechanics and Physics ofÊSolids (JMPS) 
(http://arXiv.org/abs/cond-mat/0404035)

Nelson, W., 1990. Accelerated Testing: Statistical Models, Test Plans and
Data Analyses (John Wiley \& Sons, Inc., New York).

Omori, F., 1894. On the aftershocks of earthquakes.
{\it J. Coll. Sci. Imp. Univ. Tokyo}, 7, 111.

Politi, A., Ciliberto, S., Scorretti, R., 2002. Failure time in the 
fiber-bundle model with thermal
noise and disorder. {\it Phys. Rev. E}, 66 (2), 026107/1-6.

Pradhan, S.  and Chakrabarti, B.K., 2002. Precursors of catastrophe in the
Bak-Tang-Wiesenfeld, Manna, and random-fiber-bundle models of failure,
{\it Phys. Rev. E.}, 016113.

Pradhan, S.  and Chakrabarti, B.K., 2003.
Failure due to fatigue in fiber bundles and solids,
{\it Phys. Rev. E.} 67, 046124.

Pradhan, S. and B.K. Chakrabarti, 2004.
Failure properties of fiber bundle models,
{\it Int. J. Mod. Phys. B.} 17( 29), 5565-5581. 

Ramanathan, S. and D.S. Fisher, 1998. Onset of propagation
of planar cracks in heterogenous media. {\it Phys. Rev. B.}, 58, 6026-6046.

Reichhardt, T., 1996. Rocket failure leads to grounding of small 
US satellites. {\it Nature (London)}, 384, 99-99.
 
Roux, S., Hansen, A., Herrmann, H., Guyon, E., 1988.
Rupture of heterogeneous media in the limit of infinite disorder.
{\it J. Stat. Phys.}, 52, 237-244.

Sahimi, M. and S. Arbabi, 1996. Scaling laws for fracture of
heterogeneous materials and rock. {\it Phys. Rev. Lett.}, 77, 3689-3692.

Saichev, A., Sornette, D., 2005. Andrade, 
Omori and time-to-failure laws from thermal noise in
material rupture, Phys. Rev. E 71 (1)
preprint http://arXiv.org/abs/cond-mat/0311493.

Sammis, S.G. and D. Sornette, 2002.
Positive Feedback, Memory and the Predictability of Earthquakes.
{\it Proc. Nat. Acad. Sci. USA}, 99 (SUPP1), 2501-2508.

Shcherbakov, R., Turcotte, D.L., 2003. Damage and 
self-similarity in fracture. {\it Theoretical and
Applied Fracture Mechanics}, 39 (3), 245-258.

Sornette, D., 1998.
Discrete scale invariance and complex dimensions, 
{\it Physics Reports}, 297, 239-270.

Sornette, D., 2002.
Predictability of catastrophic events: material rupture, earthquakes, turbulence, 
financial crashes and human birth. {\it Proc. Nat. Acad. 
Sci. USA}, 99 (SUPP1), 2522-2529.

Sornette, D. and  J. V. Andersen, 1998.
Scaling with respect to disorder in time-to-failure.
{\it Eur. Phys. Journal B}, 1, 353-357.

Sornette, D. and C. Vanneste, 1992.
Dynamics and memory effects in rupture of thermal fuse networks.
{\it Phys. Rev. Lett.}, 68, 612-615.

Sornette, D. and Vanneste C., 1994.
Dendrites and fronts in a model of dynamical rupture with
damage. {\it Phys. Rev. E}, 50, 4327-4345.

Stauffer D. and A. Aharony, 1992. Percolation theory (Taylor and
Francis, London).

Tang, C.A., H. Liu, P.K.K. Lee, Y. Tsui and L.G. Tham, 2000.
Numerical studies of the infuence of microstructure on rock
failure in uniaxial compression -- Part I: effect of heterogeneity.
{\it Int. J. of Rock Mechanics and Mining Sciences}, 37, 555-569.

Tang, C.A., H. Liu, P.K.K. Lee, Y. Tsui and L.G. Tham, 2000.
Numerical studies of the infuence of microstructure on rock
failure in uniaxial compression -- Part II: constraint, slenderness
and size effect.
{\it Int. J. of Rock Mechanics and Mining Sciences}, 37, 571-583.

Turcotte, D.L., Newman W.I., Shcherbakov, R., 2003. Micro and macroscopic models of rock
fracture. {\it Geophys. J. Int.}, 152 (3), 718-728.

von Neumann, J. and O. Morgenstern, 1947.
Theory of games and economic behavior.
Princetown University Press.

Vujosevic, M., Krajcinovic, D., 1997. Creep rupture of polymers -- 
A statistical model. {\it Int. J Solid
Structures}, 34 (9), 1105-1122.

Westwood, A. R. C., J. S. Ahearn and J. J. Mills, 1981. Developments in the theory
and application of chemomechanical effects. {\it Colloids and Surfaces}, 2, 1.

Yukalov, V.I., Moura A., Nechad, H., 2004. Self-similar law 
of energy release before materials
fracture. {\it J. Mech. Phys. Solids}, 52, 453-465.

Zhou, W.-X. and D. Sornette, 2002.
Generalized q-Analysis of Log-Periodicity: Applications to Critical
Ruptures. {\it Phys. Rev. E}, 046111, 6604 N4 PT2:U129-U136.

\begin{flushright}
Didier Sornette\\
Institute of Geophysics and Planetary Physics 
and Department of Earth and Space Science\\
University of California, Los Angeles, California, USA\\
and CNRS and Universit\'e des Sciences\\
Nice, France\\
sornette@moho.ess.ucla.edu
\end{flushright}

\clearpage

\clearpage

\end{document}